\documentstyle[prd,aps,floats]{revtex}

\begin{document}

\draft
\input epsf
\twocolumn[\hsize\textwidth\columnwidth\hsize\csname
@twocolumnfalse\endcsname

\title{Q-ball Formation through Affleck-Dine Mechanism}

\author{S. Kasuya$^{1}$ and M. Kawasaki$^{2}$}
\address{${}^1$ Department of Physics, Ochanomizu University,
  Bunkyo-ku, Tokyo 112-8610, Japan}
\address{${}^2$ Research Center for the Early Universe, University 
  of Tokyo, Bunkyo-ku, Tokyo 113-0033, Japan}

\date{September 25, 1999}

\maketitle

\begin{abstract}
We present the full nonlinear calculation of the formation of
a Q-ball through the Affleck-Dine (AD) mechanism by numerical
simulations. It is shown that large Q-balls are actually produced by
the fragmentation of the condensate of a scalar field whose potential
is very flat. We find that the typical size of a Q-ball is determined
by the most developed mode of linearized fluctuations, and
almost all the initial charges which the AD condensate carries are
absorbed into the formed Q-balls, whose sizes and the charges depend
only on the initial charge densities.

\end{abstract}

\pacs{PACS numbers: 98.80.Cq, 11.27.+d, 11.30.Fs
      \hspace{5cm} hep-ph/9909509}


\vskip2pc]

\setcounter{footnote}{1}
\renewcommand{\thefootnote}{\fnsymbol{footnote}}

A Q-ball is a non-topological soliton with some conserved global
charge in the scalar field theory \cite{Coleman}. A Q-ball solution
exists if the energy minimum develops at non-zero $\phi$ with fixed 
charge $Q$. In terms of the effective potential of the field $\phi$,
$V(\phi)/\phi^2$ takes the minimum at $\phi \neq 0$ 
\cite{Coleman,Kusenko1}. The Q-ball naturally appears in the spectrum
of the minimal supersymmetric standard model (MSSM)
\cite{Kusenko2}. In particular, very large Q-balls could exist in the
theory with a very flat potential \cite{Dvali}, such as in the MSSM
which has many flat directions, which consist of squarks and sleptons, 
carrying baryonic and/or leptonic charges \cite{Dine}. It
provides interesting attention to phenomenology and astrophysics
\cite{Dvali,KSTT,KKST}.

Cosmologically, it is interesting that large Q-balls carrying baryonic
charge (B-ball) can be promising candidates for the dark matter of the
universe, and/or the source for the baryogenesis
\cite{KuSh,EnMc1,EnMc2}. Moreover, they can explain why the energy
density of baryons is as large as that of the dark matter (at least
within a few orders of magnitude). If the effective potential of the
field $\phi$ carrying the baryonic charge is very flat at large
$\phi$, as in the theory that the supersymmetry (SUSY) breaking occurs
at low energy scales (gauge-mediated SUSY breaking), B-ball energy per 
unit charge decreases as the charge increases \cite{Dvali}. For large
enough charge, such as $B \gtrsim 10^{12}$, the B-ball cannot decay
into nucleon, and is completely stable, which implies that B-balls
themselves can be the dark matter \cite{KuSh} with charges
$B=10^{14}-10^{26}$ \cite{LaSh}, while baryons are created by the
conventional Affleck-Dine (AD) mechanism. In the case of
gravity-mediated SUSY breaking scenario, the B-balls can decay into
quarks or nucleons, with the decay (evaporation) rate of the Q-ball
proportional to the surface area \cite{Cohen}, and if they decay after
the electroweak phase transition, there are some advantages over the
conventional AD baryogenesis \cite{EnMc1,EnMc2}. For example, B-balls
can protect the baryon asymmetry from the effects of lepton violating
interactions above the electroweak scale when anomalous $B+L$
violation is in thermal equilibrium. Another one is that Q-balls with
$B-L$ charge survive the sphaleron effects to create the same amounts
of baryon and lepton numbers. In either case, it is necessary for
Q-balls to have large charges, such as $Q=10^{22}-10^{28}$
\cite{EnMc1,EnMc2}. In this scenario, dark matter is lightest
supersymmetric particle (LSP) which arises from the Q-ball decay, and
parameters of MSSM could be constrained by investigating the Q-ball
cosmology \cite{EnMc3}. Note that it is also possible to have stable
Q-balls in the gravity-mediated SUSY breaking theory depending on the
details of the features of the hidden sector \cite{EnMc4}. 
   
Those large Q-balls are expected to be created through AD mechanism
\cite{AfDi} in the inflationary universe \cite{KuSh,EnMc1,EnMc2}. The  
coherent state of the AD scalar field which consists of some flat
direction in MSSM becomes unstable and instabilities develop. These
fluctuations grow large to form Q-balls. 

The formation of large Q-balls has been studied analytically only with 
linear theory \cite{KuSh,EnMc1,EnMc2} and numerical simulations were
done in one dimensional lattices \cite{KuSh}. Both of them are based
on the assumption that the Q-ball configuration is spherical so that
we cannot really tell that the Q-ball configuration is actually
accomplished. Recently, some aspects of the dynamics of AD scalar and 
Q-ball formation were studied in Ref.~\cite{EnMc5}, but the whole
evolution was not investigated, which is important for the
investigation of the Q-ball formation. (In the context of
non-relativistic Bose gas, the dynamics of drops of Bose-Einstein
condensate, which is a non-topological soliton, are studied in
Ref.~\cite{KhTk1}.) In this Rapid Communication, we study the dynamics
of a complex scalar field with very flat potential numerically in one,
two, and three dimensional lattices, without assuming spherical Q-ball
configuration. On one dimensional lattices, it is equivalent to the
system independent of other two dimensions, so that we are observing
plane-like objects. We call them {\it Q-walls}. Likewise, string-like
objects, which we call {\it Q-strings}, appear on two dimensional
lattices. 

First we show where the instabilities of a scalar field come from. To
be concrete, let us assume that the complex scalar field $\Phi$ has an
effective potential of the form~\cite{KuSh}
\begin{equation}
  \label{pot}
  V(\Phi) = m^4\ln\left( 1+\frac{|\Phi|^2}{m^2} \right) 
           - cH^2|\Phi|^2 + \frac{\lambda^2}{M^2}|\Phi|^6,
\end{equation}
where $m$ is the mass of the field, $H$ is the Hubble parameter,
$\lambda$ is a dimensionless coupling constant, 
$M\simeq2.4\times 10^{18}$ GeV is the (reduced) Planck mass, and $c$
is a constant. Hereafter, we assume the matter-dominated
universe, where $H=2/3t$. This form of the potential arises naturally
in the gauge-mediated SUSY breaking scenario in MSSM~\cite{KuSh}. In
addition to Eq.(\ref{pot}), the A-term (e.g., $A(\Phi^4+\Phi^{*4})$)
is necessary for baryogenesis, since it makes the AD field rotate to
create baryon numbers. Here we assume that the A-term does not
crucially affect other terms of the potential, and take {\it ad hoc}
initial conditions neglecting the A-term. The AD field thus has the
initial charge density. Notice that the effective potential for the
flat direction has very similar form in the gravity-mediated SUSY
breaking scenario \cite{EnMc1}, and the features of the  Q-ball
formation are expected to be the same.

If we note that the field is dominated by the logarithmic potential
and the homogeneous mode has only a real part (i.e., no rotational
motion), neglecting the second and third term in Eq.(\ref{pot}),
the instability band is approximately estimated as~\cite{KK}
\begin{equation}
  \frac{k}{a}\lesssim \frac{m^2}{|\Phi|},
\end{equation}
for large field value $|\Phi|$, which is exactly the same as the
result of Ref.~\cite{KuSh} (note that there are additional instability 
bands which come from the parametric resonance effects because of the
oscillation of the homogeneous field, but they are subdominant
effects). Therefore, the instability band grows as time goes on, since 
the amplitude of the homogeneous mode $|\Phi|$ decreases as $a^{-3}$ 
for the logarithmic potential. These fluctuations originate from
the negative curvature of the logarithmic potential, which produces
negative pressure \cite{EnMc1}. It is expected that Q-balls with 
corresponding scales are formed.

As we mentioned, we calculate the dynamics of the complex scalar field
on one, two, and three dimensional lattices. We formulate the equation
of motion for the real and imaginary part of the field: 
$\Phi=\frac{1}{\sqrt{2}}(\phi_1+i\phi_2)$. Rescaling variables with
respect to $m$, we have dimensionless variables 
\begin{equation}
\varphi=\frac{\phi}{m}, \qquad h=\frac{H}{m}, \qquad
\tau = mt, \qquad \xi = mx.
\end{equation}

\begin{figure}[t!]
\centering
\hspace*{-7mm}
\leavevmode\epsfysize=8cm \epsfbox{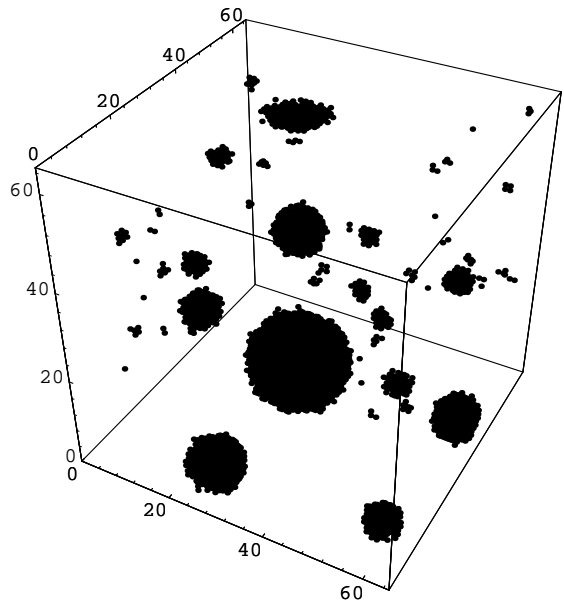}\\[2mm]
\caption[fig-1]{\label{fig-1} 
Configuration of Q-balls on three dimensional lattice. More than 30
Q-balls are formed, and the largest one has the charge with 
$Q\simeq 1.96\times 10^{16}$}
\end{figure}

We have exhaustedly calculated the dynamics of the scalar field and
Q-balls for various parameters, and find that Q-balls are actually
formed through the Affleck-Dine mechanism in three dimensional
lattices. They have thick-wall profiles which are approximately
spherical, and its charge is conserved as time goes on. Here we show
one example. Figure~\ref{fig-1} shows the configuration of the Q-ball
at $\tau=10^6$ on $64^3$ three dimensional lattices with 
$\Delta\xi = 1.0$ in the matter dominated universe. Initial conditions
are $\varphi_1(0)=2.5\times 10^6, \varphi_1'(0)=0, \varphi_2(0)=0, 
\varphi_2'(0)=4.0\times 10^4, \tau(0) = 100$. We can see more than 30
Q-balls in the box. The charge of the largest Q-ball is 
$Q \simeq 1.96\times 10^{16}$ evaluated
by 
\begin{equation}
\label{charge}
Q=\int d^3\xi q 
= \int d^3\xi \frac{1}{2}(\varphi_1\varphi_2'-\varphi_2\varphi_1').
\end{equation}

Though the best way to investigate the nature of Q-ball formation is
to calculate in three dimensions, we must use the large-sized box to
take into account low momenta effects and the large number of lattices
in order to have the enough resolution. Therefore, we have calculated
in one and two dimensions, and use these results complementarily with
results in three dimensions.

Therefore, let us now compare the evolutions of Q-balls in one, two,
and three dimensions. As is mentioned, a Q-ball is a non-trivial
configuration of the scalar field, which we can obtain by estimating
the energy minimum with finite charge $Q$ fixed. From
Eq.(\ref{charge}) we can approximately estimate the charge of a Q-ball
as 
\begin{equation}
  \label{q-conserv}
  Q = a^3 Q_D \sim a^3 R^D \tilde{q} \sim {\rm const.},
\end{equation}
where $\tilde{q}=\phi_1\dot{\phi_2}-\dot{\phi_1}\phi_2$ is the charge
density, and $Q_D$ is the charge in $D$ dimension. Charge conservation 
tells us that $Q$ is constant. If we assume the form of a Q-ball as
\begin{equation}
  \phi({\bf x},t)=\phi({\bf x})\exp({i\omega t}),
\end{equation}
the energy of a Q-ball can be calculated as
\begin{eqnarray}
  E & = & \int d^3x \left[ \frac{1}{2}(\nabla\phi)^2 + V(\phi)
        -\frac{1}{2}\omega^2\phi^2 \right] +\omega Q \nonumber \\
    & = & \int d^3x [ E_{grad} + V_1 + V_2 ] + \omega Q,
\end{eqnarray}
where 
\begin{eqnarray}
  E_{grad} & \sim & \frac{\phi^2}{a^2R^2}, \nonumber \\
  V_1      & \sim & m^4 \log\left(1+\frac{\phi^2}{2m^2}\right) 
                    \sim {\rm const.}, \nonumber \\
  V_2      & \sim & \omega^2\phi^2.
\end{eqnarray}

When the energy takes the minimum value, the equipartition is
achieved: $E_{grad} \sim V_1$ and $E_{grad} \sim V_2$. From these
equations and the charge conservation, we obtain the following 
evolutions:
\begin{eqnarray}
  R       & \propto & a^{-4/(D+1)}, \nonumber \\
  \phi    & \propto & a^{(D-3)/(D+1)}, \nonumber \\
  \omega  & \propto & a^{-(D-3)/(D+1)},
\end{eqnarray}
which we observed have approximately the same features numerically
\cite{KK}. Therefore, the physical size $R_{phys}=Ra$ of the Q-balls
for $D=1,2$ shrinks, while remaining constant for $D=3$. Thus, stable
Q-balls (three dimensional objects) can be formed, but Q-walls or
Q-strings shrink because of the charge conservation.

\begin{figure}[t!]
\centering
\hspace*{-7mm}
\leavevmode\epsfysize=8cm \epsfbox{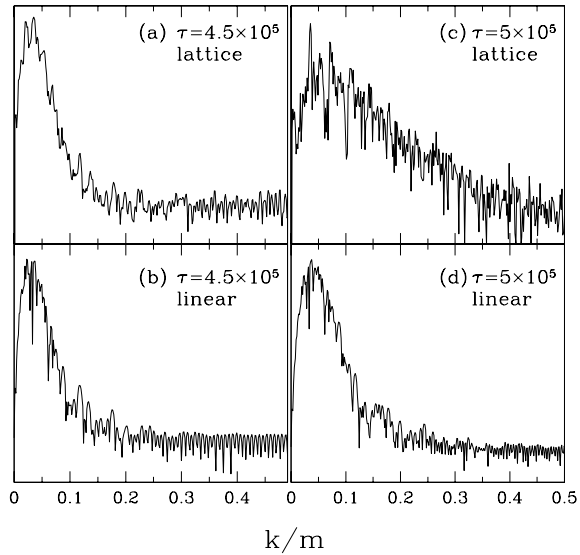}\\[2mm]
\caption[fig-2]{\label{fig-2} 
Power spectrum of fluctuations of AD scalar field when the amplitude
of fluctuations becomes as large as that of the homogeneous mode:
$\langle \delta \phi^2 \rangle \sim \phi^2$. The top panels (a) and
(c) show the full fluctuations calculated on one dimensional lattices,
while the bottom panels (b) and (d) show the linearized fluctuations
without mode mixing.} 
\end{figure}

Figure~\ref{fig-2} shows the power spectra of the cases of one
dimensional lattices with the box size $L=4096$ and linearized
fluctuations at $\tau=4.5\times 10^5$ and $5\times 10^5$. As expected,
both spectra are very similar at $\tau=4.5\times 10^5$, since
fluctuations have not fully developed yet. After they are fully
developed ($\tau=5\times 10^5$), the spectrum becomes smooth and broad
because of rescattering \cite{KhTk2} (panel 2(c)). But, even at this
time, the most developed mode is the same as that of linearized
fluctuations (Compare panels 2(c) and 2(d)). Therefore, we conclude
that the size of Q-balls is determined by the most developed mode of
linearized fluctuations when the amplitude of fluctuations becomes as
large as that of the homogeneous mode,
$\langle \delta \phi^2 \rangle \sim \phi^2$. For the case of
Fig.~\ref{fig-2}, the typical size is $k_{max} \sim 0.04$, 
which implies that 
$R_{phys} \sim a(\tau_f)/k_{max} \sim 7.3 \times 10^3$, where 
$\tau_f = 5\times 10^5$ is the Q-ball formation time. At this time,
the ratio to the horizon size is $\sim 10^{-2}$, which corresponds to
the results of Ref.~\cite{KuSh}, but this value may not have important 
meaning, since the sizes of the horizon and Q-balls have different
time evolutions. This size is consistent with the actual sizes
appearing on three dimensional lattices, where the largest Q-ball has
the size $R_{phys} \approx 1.1\times 10^4$, and the average size of
the second to fifth largest Q-balls is $\simeq 5.2\times 10^3$. 

We also observed that almost all the initial charges carried by the
condensate of AD field are absorbed into Q-balls formed from the
fragmentations of the condensate, and the amplitude of the homogeneous
mode is highly damped, which means that they carry only a small
fraction of the total initial charges. In the case of 
Fig.~\ref{fig-1}, more than $95 \%$ of the charges are stored in the 
Q-balls. 

\begin{figure}[t!]
\centering
\hspace*{-7mm}
\leavevmode\epsfysize=8cm \epsfbox{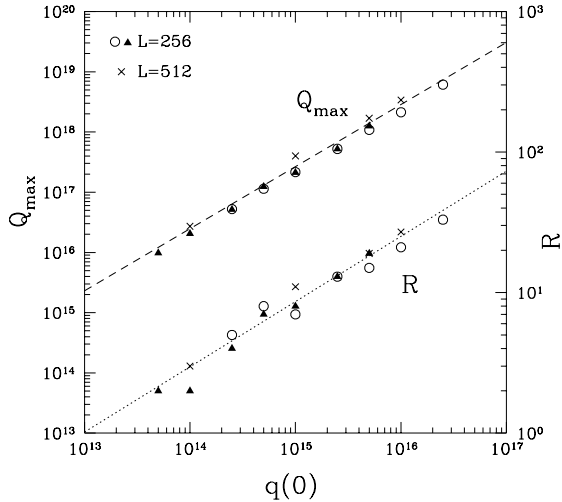}\\[2mm]
\caption[fig-3]{\label{fig-3} 
Dependence of charges and sizes on the initial charge 
$q(0)=\varphi_1(0)\varphi_2'(0)$ carried by the AD condensate on one
dimensional lattices. Open circles and solid triangles denote the
dependence on $\varphi_2'(0)$ and $\varphi_1(0)$, respectively. 
Crosses are obtained on larger box size. $R$ is evaluated at $\tau=5
\times 10^7$. Note that the relevant variable is only $q(0)$, since
both dependences lie on the same lines. } 
\end{figure}

Actually, the charges and sizes of Q-balls depend on the initial
value of the charge carried by the AD condensate. Since the initial
charge density of AD scalar is written as 
$q(0)=\varphi_1(0)\varphi_2'(0)$, the larger the initial amplitude or
angular velocity of the AD condensate, the larger the charge stored in
Q-balls. The dependences of the charge and the (comoving) size of the
(largest) Q-ball on the initial charge density are shown in 
Fig.~\ref{fig-3}. Here we take one dimensional lattices, so actually
we observe Q-walls. We thus estimate their conserved charges as 
$Q_{max}=a^3\int dx q(x)$ according to Eq.(\ref{q-conserv}). We find
that one large Q-ball and a few small Q-balls are formed in most 
of the cases, while one Q-ball is formed in some cases such that the
initial charge density $q(0)$ is relatively small. Open circles denote
the dependence on $\varphi_2'(0)$ for fixed $\varphi_1(0)$, while
solid triangles denote the dependence on $\varphi_1(0)$ for fixed
$\varphi_2'(0)$ with $L=256$. Since both results show the same
dependence, the only relevant variable which determines the charge and
the size is the initial charge density $q(0)$. The dashed line
represents the fitted line, $Q_{max}=94.3 (q(0))^{1.03}$. Crosses are
obtained on $L=512$ lattices. They seem to be a little larger, and the
box size effect might be remained. Notice that those Q-balls with
negative charge can be produced when the initial angular velocity of
the AD condensate (or, in our simulations, $\varphi_2'(0)$) is small
enough \cite{KK}. 

Since the (comoving) size of the Q-balls changes as time goes on, as
we mentioned above, the actual values of the size at 
$\tau=5\times 10^7$ in one dimension is not so important. What is
important is how the size depends on $q(0)$. The dotted line shows the
fitted line written as $R=9.46\times10^{-7}(q(0))^{0.464}$, though we
expect the relation $R \propto (q(0))^{1/3}$, which means that the
charge is proportional to the volume as in Eq.(\ref{q-conserv}). It
may be one of the reasons for the discrepancy that the values of $R$
for small charge Q-balls may have considerable error because of poor
resolution in spatial lattices.

In conclusion, we consider the full nonlinear equations of motion of 
the Affleck-Dine scalar field in order to see the formation of the
Q-ball through the Affleck-Dine mechanism by numerical simulations. It
is shown that large Q-balls are actually produced by the fragmentation
of the condensate of a scalar field whose potential is very flat, as
in the supersymmetric standard theory. 

We find that the typical size of Q-balls is determined by that of the
most developed mode of linearized fluctuations when the amplitude of
fluctuations grows as large as that of the homogeneous mode: 
$\langle \delta \phi^2 \rangle \sim \phi^2$. Almost all the
initial charges carried by the AD condensate are absorbed into the
Q-balls formed, leaving only a small fraction in the form of the
remaining coherently oscillating AD condensate. We thus can constrain
parameters of MSSM through the fraction of the baryon in Q-balls
($f_B$) in the context of the Q-ball decay producing LSP dark matter
\cite{EnMc3} 

Moreover, the actual sizes and the charges stored within Q-balls
depend on the initial charge density of the AD condensate, which is in
good agreement with the condition of the existence of the Q-ball; that
is, it exists if the scalar field can take non-trivial energy minimum
configuration with a fixed charge. Therefore, Q-balls with huge
charges necessary for Q-balls to be dark matter could be produced if
the initial charge density that the AD condensate carries is large
enough. 

We also find that the evolution of Q-balls crucially
depends on the form of their dimensions, and the stable Q-balls can
only exist in the form of three dimensional objects. Smaller
dimensional objects such as Q-walls and Q-strings shrink as the
universe expands. 

Finally, we will mention the Q-axiton (the higher energy state
Q-ball), which was studied in Ref.~\cite{EnMc5}. In our simulations,
Q-balls are actually formed. The field orbits in the complex plane
inside Q-balls are almost complete circles. Moreover, even if the
initial AD field orbit is extremely oblique such that the initial
angular velocity is very small so as to create the negatively charged 
Q-balls, circular orbits can be seen inside both positive and negative
Q-balls \cite{KK}. It thus seems that Q-axitons may appear, if ever,
at the very beginning of the Q-ball formation.

M.K. is supported in part by the Grant-in-Aid, Priority
Area ``Supersymmetry and Unified Theory of Elementary
Particles''($\#707$).

\end{document}